\documentclass[12pt,letterpaper]{article}
\usepackage{epsfig,graphicx,bm,amsmath}
\usepackage{amssymb}
\numberwithin{equation}{section}

\newcommand{\mc}[1]{\mathcal{#1}}

\begin{document}
\begin{flushright}
AEI-2010-039
\end{flushright}

\setlength{\unitlength}{1mm}

\thispagestyle{empty}
\vspace*{3cm}

\begin{center}
{\bf \Large Thermodynamic instability of doubly spinning black objects}\\
\vspace*{2cm}

{\bf Dumitru Astefanesei,}\footnote{E-mail: {\tt dumitru@aei.mpg.de}}
{\bf Maria J. Rodriguez,}\footnote{E-mail: {\tt maria.rodriguez@aei.mpg.de}}
{\bf and Stefan Theisen}\footnote{E-mail: {\tt theisen@aei.mpg.de}}

\vspace*{0.2cm}

{\it Max-Planck-Institut f\"ur Gravitationsphysik,}\\
{\it Albert-Einstein-Institut, 14476 Golm, Germany}\\[.5em]

\vspace{2cm} {\bf ABSTRACT} 
\end{center}

We investigate the thermodynamic stability of neutral black objects 
with (at least) two angular momenta. We use the quasilocal formalism 
to compute the grand canonical potential and show that the doubly 
spinning black ring is thermodynamically unstable. We consider the 
thermodynamic instabilities of ultra-spinning black objects and point 
out a subtle relation between the microcanonical and grand canonical 
ensembles. We also find the location of the black string/membrane phases 
of doubly spinning black objects.

\vfill \setcounter{page}{0} \setcounter{footnote}{0}
\newpage

\tableofcontents
\section{Introduction} \label{intro}

The physics of event horizons in higher-dimensional General Relativity (GR) 
is an interesting area of research not just for its intrinsic relevance 
to string theory. An investigation of black hole solutions in higher 
dimensions is also important because it has revealed new features:
a richer rotation dynamics and the existence of 
regular extended black hole solutions.\\
\indent It is clear by now that some of the remarkable properties of 
four-dimensional black holes do not hold in general. A notorious 
example of particular importance concerns their horizon topology. 
In four dimensions, the spherical $S^2$ topology is the only 
allowed horizon topology for asymptotically flat (AF) stationary black holes. 
A related result is the `uniqueness theorem', which states that a 
stationary AF vacuum 
black hole in four dimensions is characterized by 
its mass and angular momentum and has no other independent 
characteristic (hair).\\
\indent The spectrum of stationary black objects is far richer in dimensions bigger 
than four (see \cite{Rodriguez:2010zw} for a concise review). The restrictions on the topology of AF black holes require that spatial sections of the event horizon must be positive Yamabe type \cite{Galloway:2005mf} and if spinning (stationary) they have to be axisymmetric \cite{Hollands:2006rj}. The most obvious indication
is the existence of a neutral AF solution describing a 
spinning black ring in five dimensions \cite{Emparan:2001wn,Pomeransky:2006bd}. 
As it was shown in \cite{Emparan:2001wn}, there are
{\it three} solutions with the same asymptotic conserved 
charges (the same mass and angular momentum). On top of the well known Myers-Perry (MP) 
black hole \cite{Myers:1986un} with an $S^3$ horizon  there are two different black 
rings with an $S^1\times S^2$ horizon ($S^1\times S^{D-3}$ in $D$ dimensions). Therefore, unlike in four dimensions, 
the black objects in higher dimensions are not completely determined by a few conserved
asymptotic charges.\footnote{In \cite{Astefanesei:2009wi} it was proposed 
that the necessary information to distinguish between black objects 
with different horizon topologies is encoded in the subleading terms 
in the boundary stress tensor.} But the space of solutions of Einstein's equations in $D$ dimensions also includes extended black holes. 
The black p-branes \cite{Horowitz:1991cd}, dubbed black strings when $p=1$ or otherwise membranes, are transverse asymptotically flat (only AF in $D-p$ directions),
evade the no-hair theorems, and exhibit horizon topologies $S^{D-2-p} \times \mathbb{R}^p$. Interestingly enough, due to the richer rotational dynamics, in certain regimes in higher dimensions the thermodynamical properties of the compact black holes resemble those of the extended black objects. In this paper we aim to make progress towards a better understanding on these properties of black objects.\\
\indent We first study in detail the thermodynamic instabilities of doubly spinning 
black objects. We identify the ultra-spinning regimes (in parameter space) from 
their thermodynamic quantities. Our motivation 
stems from the observation made in \cite{Emparan:2003sy} that, 
in dimensions greater than four, the 
thermodynamics of certain fast spinning MP black holes show a qualitative 
change in behaviour. That is, there is a transition towards a black 
membrane-like behaviour. 
This is due to the fact that, as one increases the angular 
momentum, the temperature of these ultra-spinning black holes reaches a minimum and then starts to grow as expected for the black membrane.\footnote{The numerical evidence of
\cite{Dias:2009iu} suggests that the onset of the ultra-spinning regime for singly spinning MP black holes corresponds to a zero mode associated to the Gregory-Laflamme instability \cite{Gregory:1993vy}.}\\
\indent Black rings also exhibit a change in the thermodynamic behaviour that resembles the ultra-spinning regime of black holes when ultra-spinning along the $S^1$ direction.
In the `thin ring approximation', when the radius of $S^1$ is much larger 
than the radius of $S^{D-3}$, the singly spinning black ring can be approximated by a 
boosted black string. An interesting question 
we would like to address in this paper is how the ultra-spinning 
regime of the $D=5$ black ring is affected by adding the second angular momentum, along the $S^2$.
In other words, whether neutral doubly spinning black ring is thermodynamically stable in the grand canonical ensemble and whether there is any connection with its ultra-spinning regime.\\
\indent Since there is no known background subtraction method for these black
objects, we use a slightly modified version of the quasilocal 
formalism of Brown and York \cite{Brown:1992br} to address these questions. 
Supplemented with counterterms \cite{Lau:1999dp}, 
the quasilocal formalism becomes a very powerful tool to study 
the thermodynamics of black objects that are AF. Recently, several concrete five-dimensional examples were discussed in 
detail in \cite{Astefanesei:2009wi,Kleihaus:2006ee} and in \cite{Astefanesei:2009mc} for the unbalanced black ring.\\
\indent The relevant thermodynamic potential in the grand canonical ensemble is
the Gibbs potential that is the Euclidean action divided by the 
periodicity of the Euclidean time \cite{Gibbons:1976ue}.  
The Euclidean method was applied to the black ring thermodynamics in 
\cite{Astefanesei:2005ad}. Since the black ring does not have a real 
non-singular Euclidean section, the `quasi-Euclidean' method \cite{Brown:1990fk} 
was adopted to analyze the black ring thermodynamics. In this approach, 
the horizon is described by the `bolt' in a {\it complexified} Euclidean 
geometry rather than a real one.\footnote{The complex geometry is obtained by the usual 
analytic continuation of time coordinate, $\tau=it$.} It was also pointed out in \cite{Astefanesei:2005ad} that the neutral black ring with one angular momentum is unstable to angular fluctuations --- a more 
detailed analysis can be found in \cite{Astefanesei:2009wi,Elvang:2006dd, Monteiro:2009tc}.\\
\indent By employing this method to compute the Gibbs potential, we find the response functions directly in the grand canonical ensemble. We observe that the second angular momentum changes the situation in the sense that, unlike the black ring with one angular momentum, the doubly spinning 
ring is stable against perturbations in the angular velocity in some 
specific region of the parameter space.\footnote{For a black ring 
with one angular momentum the `isothermal compressibility' (moment 
of inertia) is always negative \cite{Astefanesei:2009wi}.} However, a careful analysis 
of all response functions that characterize the system reveals that 
the doubly spinning black ring is thermodynamically {\it unstable} in 
the grand canonical ensemble. That is, there is no region 
in the parameter space in which all response functions are positive 
definite.\\
\indent On general physical grounds it is expected that the microcanonical ensemble 
of asymptotically flat black holes is dominated by diffuse radiation states 
rather than black hole states \cite{Page:2004xp}. In other words, 
it is favorable for the black hole to decay away (the heat capacity is 
negative --- see Appendix B for a detailed discussion on local thermodynamic 
conditions) and so pure thermal radiation is a local equilibrium state.
Indeed, it was shown in \cite{Dias:2010eu} that, for any vacuum black hole 
characterized by its mass and angular momenta, the Hessian of the entropy 
always has a negative eigenvalue. Since the Hessian of the entropy is related 
to the inverse of the Hessian of the Gibbs potential \cite{Ruppeiner:1995zz}, 
this implies at generic points in the moduli space, i.e. away from the hypersurfaces 
defined by a vanishing eigenvalue, an instability in the grand canonical 
ensemble \cite{Dias:2010eu}.\\
\indent Identifying which doubly spinning black hole solutions exhibit this ultra-spinning regime
will be our next objective. We will first find the threshold of the black membrane regime 
for MP black holes with at least two large angular momenta. To compare the ultra-spinning regimes 
of both, the black hole and black ring, we present a careful analysis of their ultra-spinning regime.
For the black ring we observe a similar but slightly different ultra-spinning phase. 
That is, the temperature does not have a minimum but rather there is a `turning point', 
which is responsible for the boosted black string behaviour. We also find that even after adding 
the second angular momentum the ring can have a membrane-like behavior. However, 
for large enough values of the second angular momentum the `membrane phase' disappears.\\
\indent The dynamical instabilities were related to the thermodynamic ones by 
(a conjecture of) Gubser and Mitra \cite{Gubser:2000mm}: gravitational backgrounds with 
translationally invariant horizon develop a tachyonic mode (negative 
non-conformal mode) whenever the specific heat of the black brane geometry 
becomes negative. Our goal is not to study dynamical instabilities, which would
require a perturbative analysis, but rather to generalize the arguments of 
\cite{Emparan:2003sy} to black rings. However, to connect the work 
of \cite{Emparan:2003sy} to \cite{Gubser:2000mm}, one has to understand the black 
hole thermodynamics in the grand canonical ensemble (see, e.g., \cite{Reall:2001ag}). 
At first sight, it is surprising that an analysis in the microcanonical ensemble 
\cite{Emparan:2003sy} provides information about the membrane phase. 
We point out a subtle relation between the microcanonical and grand 
canonical ensembles, which is valid for the analytic solutions we are 
interested in, but it may be more general, and argue that this is why 
the microcanonical ensemble also encodes information about the membrane 
phase. 

The remainder of this paper is organized as follows. We start 
in Section 2 with a brief review of the counterterm method for 
asymptotically  flat spacetimes. We then compute the stress tensor 
and the corresponding asymptotic charges for black objects with two 
angular momenta: Myers-Perry black hole, black ring, and black 
branes. In Section 3, we present an analysis of the thermodynamic 
stability of the doubly spinning ring. We compute the thermodynamic 
action and check the quantum statistical relation. We also analyze 
in great detail the response functions. In Section 4, we investigate 
the black string/membrane phases of doubly spinning black objects. 
Finally, we conclude with a discussion of our results. In Appendix A 
we present the expressions of angular velocities and temperature for a general 
metric with two angular momenta. Appendix B contains some 
general aspects of black holes thermodynamics, the local stability 
conditions, and concrete expressions for some of the response 
functions used in Section 2.

\section{Stress tensor and conserved charges} \label{stress}
In this section we apply the counterterm method to doubly spinning 
five-dimensional vacuum solutions of Einstein gravity. We explicitly 
show how to compute the boundary stress tensor and the conserved 
charges for Myers-Perry black hole, doubly spinning black ring, and 
doubly spinning black branes.

\subsection{Quasilocal formalism and conserved charges}

To begin our considerations on thermodynamics of doubly spinning black 
objects in five dimensions, we recall the description of quasilocal 
formalism \cite{Brown:1992br} supplemented with counterterms.

To define the conserved charges we use
the divergence-free boundary stress tensor proposed in
\cite{Astefanesei:2005ad}:
\begin{eqnarray}
\label{eq:Tij}
\tau_{ij}\equiv\frac{2}{\sqrt{|h|}}\frac{\delta I}{\delta h^{ij}}=
\frac{1}{8\pi G_5}\Big( K_{ij}-h_{ij}K
-\Psi(\mathcal{R}_{ij}-\mathcal{R}\,h_{ij})-h_{ij}\square\Psi+\Psi_{;ij}\Big) 
\label{Tik}
\end{eqnarray}
where $\Psi=\sqrt{3}/\sqrt{2\mathcal{R}}$, $h_{ij}$ is the induced boundary 
metric, and $\mathcal{R}_{ij}$ is its Ricci scalar. A rigorous justification and more 
details about this proposal can be found in \cite{Mann:2005yr,Mann:2006bd,Astefanesei:2006zd}.

Here, $I$ is the \textit{renormalized} action that includes counterterms,
\begin{eqnarray}
\label{eq:action}
I&=& \frac{1}{16\pi
G_5}\int_MR\,\sqrt{-g}\,d^5x+\frac{\epsilon}{8\pi G_5}\int_{\partial
M}\left(K-\sqrt{\frac{3\,\mathcal{R}}{2}}\right)\,\sqrt{|h|}\,d^4 x
\end{eqnarray}
$K$ is the extrinsic curvature of $\partial M$ and $\epsilon=+1(-1)$ if
$\partial M$ is timelike (spacelike). 

The boundary metric can be written locally in ADM-like
form
\begin{eqnarray}
\label{sigma}
 h_{ij}dx^i dx^j=-N^2\,dt^2+\sigma_{a b}\,(dy^a
+N^a\,dt)(dy^b +N^b\,dt)
\end{eqnarray}
where $N$ and $N^a$ are the lapse function and the shift vector
respectively and \{$y^a$\} are the intrinsic coordinates on a (closed)
hypersurface $\Sigma$. 
If the boundary geometry has an isometry generated by a
Killing vector $\xi ^{i}$, a conserved charge
\begin{eqnarray}
\label{charge}
 {\mathfrak Q}_{\xi }=\oint_{\Sigma }d^3y\,
\sqrt{\sigma}n^i\tau_{ij}\,\xi^j
\end{eqnarray}
can be associated with the hypersurface $\Sigma $ (with normal
$n^{i}$). 

\subsection{Doubly spinning solutions} \label{solutions}

\subsubsection{Black hole}
The Einstein equations in higher dimensions
have spinning black hole solutions \cite{Myers:1986un}. In 
five dimensions, the Myers-Perry black hole in Boyer-Lindquist
type coordinates is
\begin{eqnarray}
\label{Kerr-5D}
ds_{BH}^2 & = & -dt^2 +\Sigma\,\left(\frac{r^2}{\Delta}\,dr^2 + d\theta^2 \right)
+(r^2 + a^2)\,\sin^2\theta \,d\phi^2
+(r^2 + b^2)\,\cos^2\theta \,d\psi^2 \nonumber \\
&  & +\,\frac{m}{\Sigma}\, \left(dt - a\, \sin^2\theta \,d\phi
- b\,\cos^2\theta \,d\psi \right)^2
\label{metric}
\end{eqnarray}
where
\begin{eqnarray}
\Sigma=r^2+a^2 \,\cos^2\theta + b^2 \,\sin^2 \theta, \;\;\;\;\;\;
\Delta= (r^2 + a^2)(r^2 + b^2) -m \, r^2 
\end{eqnarray}
and $\,m \,$  is a parameter related to the physical
mass of the black hole, while the parameters $\,a \,$ and $\,b \,$
are associated with its two independent angular momenta. This metric 
depends only on two coordinates, $0< r < \infty$ and $0\le\theta\le\pi/2$, 
and it is independent of time, $-\infty < t < \infty$, and the azimuthal 
angles, $0 < \phi,\psi < 2\pi$.

Since $r$ is playing the role of a radial coordinate 
in this coordinate system, the event horizon is also the null surface determined by
the equation $\,g^{rr}=0\,$. So, the event horizon of the black hole can be computed by using (\ref{hor}), which 
implies $\Delta=0$. The largest root of this equation gives the radius of the black hole's
outer event horizon
\begin{eqnarray}
r_{h}^2 = \frac{1}{2}\,\left(m - a^2-b^2 +
\sqrt{(m - a^2- b^2)^2 - 4\,a^2\,b^2} \right)
\label{hradius}
\end{eqnarray}
Notice that the horizon exists if and only if
\begin{eqnarray}
a^2 + b^2 + 2 |a\,b| \,\leq m
\label{extreme}
\end{eqnarray}
so that the condition $ \, m=a^2 + b^2 + 2 |a\,b| \,$ or, equivalently,
$\,r_{h}^2= |a\,b|\,$  defines the extremal horizon of a
five dimensional black hole (when one angular momentum vanishes, 
the horizon area goes to zero in the extremal limit). Otherwise, the 
metric describes a naked singularity.

In the asymptotic limit, $r\rightarrow \infty$, the metric \eqref{Kerr-5D} 
approaches Minkowski space
\begin{eqnarray}
\label{FlatCoord}
ds^2=-dt^2+dr^2+r^2(d\theta^2+\sin^2\theta \, d\phi^2+\cos^2\theta \, d\psi^2)
\end{eqnarray}
We use the expression of black hole metric in Boyer-Lindquist 
coordinates to compute the boundary stress tensor and we obtain 
the following non-vanishing components:
\begin{eqnarray}
\label{Kerr-5D-Tij}
\tau_{tt}&=&\frac{1}{8\pi G_5}\left(-\frac{3}{2}\,m\, \frac{1}{r^3}-\frac{5}{3} (a^2-b^2) \frac{\cos 2 \theta}{r^3}  +\mc{O}(1/r^5)\right)\,,\nonumber\\
\tau_{t\phi}&=&\frac{1}{8\pi G_5}\left(-2 \,a\, m  \frac{\sin^2 \theta }{r^3}+\mc{O}(1/r^5)\right)\,,\nonumber\\
\tau_{t\psi}&=&\frac{1}{8\pi G_5}\left(-2\, b\, m  \frac{\cos^2 \theta}{r^3}+\mc{O}(1/r^5)\right)\,,\nonumber\\
\tau_{\theta\theta}&=&\frac{1}{8\pi G_5}\left(\frac{2}{3} \left(a^2-b^2\right) \frac{\cos 2 \theta}{r}+\mc{O}(1/r^3)\right)\,,\\
\tau_{\phi\phi}&=&\frac{1}{8\pi G_5}\left(\frac{2}{3}\left(a^2-b^2\right) \frac{(-1+2 \cos 2 \theta )\, \sin^2\theta}{ r}+\mc{O}(1/r^3)\right)\,,\nonumber\\
\tau_{\psi\psi}&=&\frac{1}{8\pi G_5}\left(\frac{2}{3} \left(a^2-b^2\right) \frac{ (1+2 \cos 2 \theta )\,\cos^2 \theta}{r}+\mc{O}(1/r^3)\right)\,,\nonumber\\
\tau_{\phi\psi}&=&\frac{1}{8\pi G_5}\left(-4 \,a\, b\, m  \frac{\cos^2\theta\, \sin^2\theta }{r^3}+\mc{O}(1/r^5)\right)\nonumber
\end{eqnarray}
This stress tensor is covariantly conserved with respect to the 
boundary metric (\ref{FlatCoord}). We also notice that, for equal 
angular momenta, the diagonal `angular' components of the stress 
tensor vanish --- this is intuitively expected due to the enhanced 
symmetry.

Using the definition (\ref{charge}), it is straightforwardly to obtain 
the conserved charges associated with the surface $\Sigma$ as
\begin{eqnarray}
M=\oint_{\Sigma} d^3y \sqrt{\sigma} n^i\,\tau_{ij}\,\xi_t^j\,,\qquad J_{\phi}=\oint_{\Sigma} d^3y 
\sqrt{\sigma} n^i\,\tau_{ij}\,\xi_{\phi}^j\,,\qquad J_{\psi}=\oint_{\Sigma} d^3y \sqrt{\sigma} n^i\,\tau_{ij}\,\xi_{\psi}^j \nonumber
\end{eqnarray}
where the normalized Killing vectors associated with the mass 
and angular momenta are $\xi_t=\partial_t$, $\xi_{\phi}=\partial_{\phi}$, 
and $\xi_{\psi}=\partial_{\psi}$ respectively.
We find  
\begin{eqnarray}
\label{charges-Kerr-5D}
M=\frac{3\, \pi\, m}{8\, G_5}\,,\qquad J_{\phi}=\frac{\pi \,m\, a}{4\, G_5}\,,\qquad J_{\psi}=\frac{\pi\, m\, b}{4\, G_5}
\end{eqnarray}
which is in perfect agreement with the $ADM$ calculation.

\subsubsection{Black ring }

A black ring is a five-dimensional black hole with an event horizon
of topology $S^1\times S^2$ and the metric was presented in 
\cite{Emparan:2001wn} --- the solution of Emparan and Reall has one 
angular momentum. In five dimensions, a more general solution for a black ring
with two angular momenta was presented by 
Pomeransky and Sen'kov \cite{Pomeransky:2006bd}. Some properties of the solution including the structure of the 
phases in the microcanonical ensemble are discussed in \cite{Elvang:2007hs}. 
A study of its geodesics has been performed in \cite{Durkee:2008an} 
and a careful investigation of global properties appeared recently in 
\cite{Chrusciel:2009vr}. We provide here a 
brief account of the doubly spinning black ring solution and compute 
the boundary stress tensor and the conserved charges. 

We will use the solution in the form presented in \cite{Pomeransky:2006bd}.
%
%
The metric depends just on the coordinates $x$ and $y$ defined within the 
following intervals $-1\leq x \leq 1$ and $-\infty < y < -1$. Notice 
that $x$ is like an angular coordinate --- this observation will be useful 
when we will define new coordinates that make asymptotic flatness clear.

%
%
%
%

The metric has a coordinate singularity where $g_{yy}$ diverges. The event 
horizon of the doubly spinning black ring is located at the smallest absolute 
value of $1+\lambda \, y+\nu \, y^2 = 0$, namely   
\begin{equation}
y_h=\frac{-\lambda+\sqrt{\lambda^2-4\nu}}{2\nu} 
\end{equation}
For a regular black ring solution, the parameters 
$\nu$ and $\lambda$ are constrained to satisfy \cite{Pomeransky:2006bd}: 
\begin{eqnarray}
\label{eq:parameterrange}
0 \le \nu < 1\,,\qquad 2\sqrt{\nu} \le \lambda < 1 +\nu 
\end{eqnarray}
In the limit $\nu\rightarrow 0$ the black ring with one angular 
momentum ($J_{\phi}$) is recovered ($J_{\psi}$ is the angular 
momentum on $S^2$). The limit $\lambda \rightarrow  2\sqrt{\nu}$ was carefully studied 
in $\cite{Elvang:2007hs}$ and shown to correspond to regular extremal black rings. 

We use a coordinate transformation similar to the one in \cite{Elvang:2007hs}:
\begin{eqnarray}
  x = -1 + 4 k^2 \, \alpha^2 \, \frac{\cos^2\theta }{r^2}  \, ,
   \hspace{0.9cm}
  y = -1 - 4 k^2 \, \alpha^2  \, \frac{\sin^2\theta }{r^2} \, ,
   \hspace{0.9cm}
  \alpha = \sqrt{\frac{1+\nu - \lambda}{1-\nu}} 
\end{eqnarray}
In these coordinates $\partial_t,\partial_\phi$, and $\partial_\psi$ 
are Killing vectors and the asymptotic metric is the same as 
\eqref{FlatCoord}.

The boundary stress tensor in these new coordinates is
\begin{eqnarray}
\label{2SpinBR-Tij}
\tau_{tt}&=&\frac{1}{8\pi G_5}\left(-\frac{12 k^2 \lambda  }{ (1+\nu-\lambda ) }\frac{1}{r^3}-\frac{8\,k^2 F_1[\nu,\lambda]}{3 (1+\nu-\lambda ) (1-\nu )^2}\frac{ \cos 2 \theta }{r^3}+\mc{O}(1/r^5)\right)\,,\nonumber\\
\tau_{t\phi}&=&\frac{1}{8\pi G_5}\left(\frac{16 k^3 \lambda  \left(1+\lambda -6 \nu +\lambda  \nu +\nu ^2\right) \sqrt{(1+\nu )^2-\lambda ^2}  }{(1 +\nu-\lambda )^2(1-\nu )^2 }\frac{\sin^2\theta }{r^3}+\mc{O}(1/r^5)\right)\,,\nonumber\\
\tau_{t\psi}&=&\frac{1}{8\pi G_5}\left(\frac{32 k^3 \lambda  \sqrt{\nu\big[(1+\nu )^2-\lambda ^2\big]} }{(1+\nu-\lambda  ) (1-\nu )^2}\frac{\cos^2\theta}{r^3}+\mc{O}(1/r^5)\right)\,,\nonumber\\
\tau_{\theta\theta}&=&\frac{1}{8\pi G_5}\left(\frac{2 k^2 F_2[\nu,\lambda] }{3 (1+\nu-\lambda ) (1-\nu )^2 }\frac{\cos 2\theta}{r}+\mc{O}(1/r^3)\right)\,,\\
\tau_{\phi\phi}&=&\frac{1}{8\pi G_5}\left(\frac{k^2 \left(-F_3[\nu,\lambda]+F_4[\nu,\lambda]\,\cos 2 \theta \right) }{3 (1+\nu-\lambda ) (1-\nu )^2 }\frac{\sin^2\theta }{r}+\mc{O}(1/r^3)\right)\,,\nonumber\\
\tau_{\psi\psi}&=&\frac{1}{8\pi G_5}\left(\frac{k^2 \left(F_3[\nu,\lambda]+F_4[\nu,\lambda]\,\cos 2 \theta \right)}{3 (1+\nu-\lambda) (1-\nu )^2 }\frac{\cos^2\theta }{r}-\mc{O}(1/r^3)\right)\,,\nonumber\\
\tau_{\phi\psi}&=&\frac{1}{8\pi G_5}\left(\frac{128 k^4 \lambda  \sqrt{\nu } \left(4 \lambda  \nu -(\lambda ^2 +(1-\nu )^2 )(1+\nu )\right) }{(1+\nu-\lambda) (1-\nu )^4}\frac{\cos^2\theta \sin^2\theta }{r^3}+\mc{O}(1/r^5)\right)\nonumber
\end{eqnarray}
where 
\begin{eqnarray}
F_1[\nu,\lambda]&=&1-5 \nu -\nu ^2+5 \nu ^3+\lambda ^2 (3+7 \nu )+\lambda  (1-14 \nu -7 \nu ^2)\,,\nonumber\\
F_2[\nu,\lambda]&=&1-11 \nu -\nu ^2+11 \nu ^3+\lambda ^2 (3+13 \nu )+4 \lambda (1-5 \nu -4 \nu ^2)\,,\nonumber\\
F_3[\nu,\lambda]&=&5-7 \nu -5 \nu ^2+7 \nu ^3+\lambda ^2 (15+17 \nu )-4 \lambda (1+13 \nu +2 \nu ^2)\,,\nonumber\\
F_4[\nu,\lambda]&=&7-29 \nu -7 \nu ^2+29 \nu ^3+\lambda ^2 (21+43 \nu )+\lambda (4-92 \nu -40 \nu ^2)\nonumber 
\end{eqnarray}

As in the case of doubly spinning black hole, this stress tensor is 
covariantly conserved with respect to the boundary metric (\ref{FlatCoord}). 
However, since for the doubly spinning black ring the angular momenta 
can not be equal, namely
\begin{eqnarray}
3 \,J_{\psi}\le J_{\phi}\,,
\end{eqnarray}
 there is no similar symmetry enhancement as in the black hole case in the angular 
part. 

By plugging the expressions of the boundary stress-energy components (\ref{2SpinBR-Tij}) 
in (\ref{charge}) we find the following expressions for the conserved charges:
\begin{eqnarray}
\label{charges-2SpinBR}
  &&M \; =\;   \frac{3 \pi\, k^2}{G_5} \frac{\lambda}{1+\nu-\lambda}\, ,
  \hspace{8mm}
  J_\psi\; =\;  
  \frac{4 \pi\, k^3}{G_5} 
  \frac{\lambda \, \sqrt{\nu \big[ (1+\nu)^2 - \lambda^2 \big]}}
         {(1+\nu-\lambda)(1-\nu)^2} \, , \\[2mm]
  &&J_\phi \; =\; 
  \frac{2 \pi\, k^3}{G_5} 
  \frac{\lambda \, (1+\lambda-6 \nu + \nu\, \lambda+\nu^2)\, \sqrt{ (1+\nu)^2 - \lambda^2}}
         {(1+\nu-\lambda)^2(1-\nu)^2} 
\end{eqnarray}

As expected, the charges computed by using the quasilocal formalism 
recover correctly the ADM results \cite{Pomeransky:2006bd}. 

In principle, one can obtain a black hole and a black ring with the 
same conserved charges. However, an asymptotic observer can not 
distinguish between a black hole and a black ring just by computing 
the conserved asymptotic charges. We would like to emphasize that 
it is expected that the subleading terms of the quasilocal stress tensor  
 encode the information necessary to distinguish between 
black objects with different horizon topologies.

\subsubsection{Black membrane}
Here we would like to apply the quasilocal formalism to doubly spinning 
black p-branes, dubbed also black membranes (BM) or black strings if $p=1$. 
The black membrane metric we are interested in is obtained 
by adding flat directions to a 5-dimensional black hole with two angular 
momenta. Therefore, the metric is
\begin{eqnarray}\label{blackbrane}
ds^2_{BM}=ds_{BH}^2+\sum_{i=1}^{D-5} dx_i^2
\end{eqnarray}
where $ds_{BH}^2$ is the black hole metric defined in (\ref{Kerr-5D}). 

Since the number of dimensions and the topology are changed, one expects 
changes with respect to the former discussion. For example, the form of 
the counterterm leading to a finite actions may be different when the 
number of dimensions is increased. However, in this particular case, 
what is important is the `seed' $5$-dimensional solution to which we 
add the flat directions. Thus, the form of the counterterm does not 
change but the stress tensor will have new components.

A similar computation as for the doubly spinning black hole reveals that 
the stress tensor of the BM is the one in  (\ref{Kerr-5D-Tij}) supplemented 
with the components in the new directions:
\begin{eqnarray}
\label{BlackBrane-Tij}
\tau_{x_i x_i}&=&\frac{1}{8\pi G_5}\left(-\frac{3}{2}\,m\, \frac{1}{r^3}-\frac{5}{3} (a^2-b^2) \frac{\cos 2 \theta}{r^3}  +\mc{O}(1/r^5)\right)
\end{eqnarray}
This result resembles the tension (per unit length) of the black string.

\section{Thermodynamic instability of the black ring}\label{thermodynamics}
In this section, we discuss the thermodynamics of a doubly spinning 
ring in the grand canonical ensemble. 

So far, we have computed the conserved charges of neutral 
spinning black objects with two angular momenta by using 
the quasilocal formalism. However, the quasilocal formalism 
is a very powerful tool for understanding the thermodynamics 
in more detail. In particular, one can compute the 
action and, therefore, the thermodynamic potential. 

In what follows,
we present a detailed analysis of thermodynamic stability 
of the doubly spinning black ring --- an analysis of the 
thermodynamic stability of Myers-Perry black hole with 
two angular momenta can be found in \cite{Monteiro:2009tc}.

Let us start by computing the angular velocities and the 
temperature for this solution. From (\ref{angvel}) we 
obtain the following expressions for the angular velocities:
\begin{eqnarray}
\label{eq:angvBR}
\Omega_{\psi}=\frac{\lambda (1+ \nu )-(1-\nu )\sqrt{\lambda ^2-4 \nu } }{4 \lambda\sqrt{\nu}k}\sqrt{\frac{1+\nu-\lambda}{1+\nu+\lambda }}\,,\qquad
\Omega_{\phi}=\frac{1}{2 k} \sqrt{\frac{1+\nu -\lambda }{1+\nu +\lambda }}
\end{eqnarray}
The area of the event horizon and the temperature (\ref{temp})
are
\begin{eqnarray}
{\cal A}_H=\frac{32  \pi^2k^3\,\lambda (1+\lambda +\nu ) }{(1-\nu )^2 (y_h^{-1}-y_h)}\,,\qquad
T=\frac{\sqrt{\lambda ^2-4 \nu } (1-\nu ) (y_h^{-1}-y_h)}{8 \pi k\, \lambda  (1+\lambda +\nu )}
\end{eqnarray}
Note that $y_h=\frac{-\lambda+\sqrt{\lambda^2-4\nu}}{2\nu}$ is the biggest root 
of (\ref{hor}) which corresponds to the outer event horizon --- at this point, it 
might be useful to emphasize again that  $-\infty < y < -1$.

The starting point of the Euclidean approach to black hole thermodynamics is 
the partition function \cite{Gibbons:1976ue}\footnote{It should be understood 
as a low energy effective theory rather than a proper theory of quantum gravity.} 
\begin{equation}
Z(\beta)=\int d[g,\phi]e^{-I[g,\phi]}
\end{equation}
where $\phi$ is a collective notation for the matter fields, 
$d[g,\phi]$ is the measure, and $I[g,\phi]$ is the Euclidean 
classical action. The gravitational partition function is defined by a sum over 
all {\it smooth} geometries (including black holes) that are periodic 
with period $\beta=T^{-1}$ in the same class of boundary conditions 
e.g., AF spacetimes.

For our purpose it is enough to consider the saddle point approximation. 
The grand canonical partition function is then 
$Z=Tre^{-\beta(H-\Omega_aJ_a)}\simeq e^{-I_{cl}}$ (here 
we are interested in black objects with two angular momenta), where 
$I_{cl}$ is the classical action. The saddle point is usually referred to 
as a {\it gravitational instanton}.\footnote{A quantum field can 
be treated as a small perturbation about the gravitational instanton. 
The next order contribution, which gives the one loop correction, 
includes also the thermal radiation outside the black hole.}

The thermodynamic (effective) potential associated to 
grand canonical ensemble is
\begin{equation}
\label{Gi}
G[T,\Omega_a]\equiv\frac{I_{cl}}{\beta}=M-TS-\Omega_a\,J_a
\end{equation}

On the Euclidean section, the topology near the horizon is 
modified\footnote{The origin in the Euclidean spacetime 
translates to the horizon surface in the Lorentzian spacetime. 
The Euclidean section can be understood as an effective 
description where the microstates can not be distinguished.} 
and one has to deal with manifolds with conical singularities. It 
was shown in \cite{Banados:1993qp, Fursaev:1995ef} that the conical defect has 
a contribution to the curvature and, consequently, the path 
integral is rescaled by $e^{S}$. However, this can be intuitively 
interpreted as a consequence of a trace over the macroscopically 
indistinguishable microstates.

Let us now compute the action for the doubly spinning black ring.
Since the Ricci scalar vanishes on-shell, the only contribution 
to the action is coming from the surface terms. To evaluate these terms, it is
convenient to use the $(r,\theta)$ coordinate system instead of the 
$(x,y)$ coordinates --- the reason is that the normal to the boundary 
has just one non-vanishing component. We find 
\begin{eqnarray}\label{eq:limitSBR} 
\lim_{r \to \infty}\sqrt{|h|}\left(\sqrt{\frac{3}{2}\mathcal{R}} -K\right)
&=&\frac{2k^2 \left(\lambda  (1-\nu )-F_5[\nu,\lambda]\cos 2 \theta \right) \sin 2 \theta }{(1+\nu-\lambda ) (1-\nu )}+{\cal O} (1/r)
\end{eqnarray}
where $F_5[\nu,\lambda]=1+3 \lambda ^2+6 \nu +5 \nu ^2-4 \lambda  -8 \lambda \nu $.
The expression for the total action is
\begin{eqnarray}\label{eq:BRaction}
I_{cl}=\beta \frac{\pi k^2 }{G_5} \frac{ \lambda } {(1+\nu-\lambda  )}
\end{eqnarray}
and satisfies (\ref{Gi}), which is the quantum statistical 
relation for the doubly spinning black ring. This can also be regarded as a 
non-trivial check that the entropy $S$ of this solution is, indeed, one quarter 
of the event horizon area ${\cal A}_H$. 

We have checked that the usual thermodynamic relations
\begin{equation}
S=-\left(\frac{\partial G}{\partial T}\right)_{\Omega_a}\,\, , \,\,\,\,\,\,\,\,\,\, 
J_a=-\left(\frac{\partial G}{\partial \Omega_a}\right)_{T, \Omega_b}
\end{equation}
are satisfied and so the Gibbs potential $G[T,\Omega_{\phi},\Omega_{\psi}]$ is indeed 
the Legendre transform of the energy $M[S,J_{\phi},J_{\psi}]$ with respect to
$S$, $J_{\phi}$, and $J_{\psi}$. 

We want to also point out that, in the light of the new developments in 
understanding the balance condition for gravity solutions 
\cite{Astefanesei:2009mc}, the form of quantum statistical relation hints 
to the fact that this solution is balanced. Indeed, our results are 
in perfect agreement with the recent detailed analysis of the global 
properties of the doubly spinning black ring \cite{Chrusciel:2009vr}.

Now, we are ready to discuss the thermodynamic stability in the grand 
canonical ensemble --- in Appendix \ref{Ap:Stability} we 
summarize the thermal stability conditions and present explicit 
expressions for some response functions we are interested in. We analyze 
in detail the response functions that signal the (in)stability of 
the black ring against fluctuations.

\begin{figure}
\centering
  \includegraphics[height=5cm,width=6.8cm]{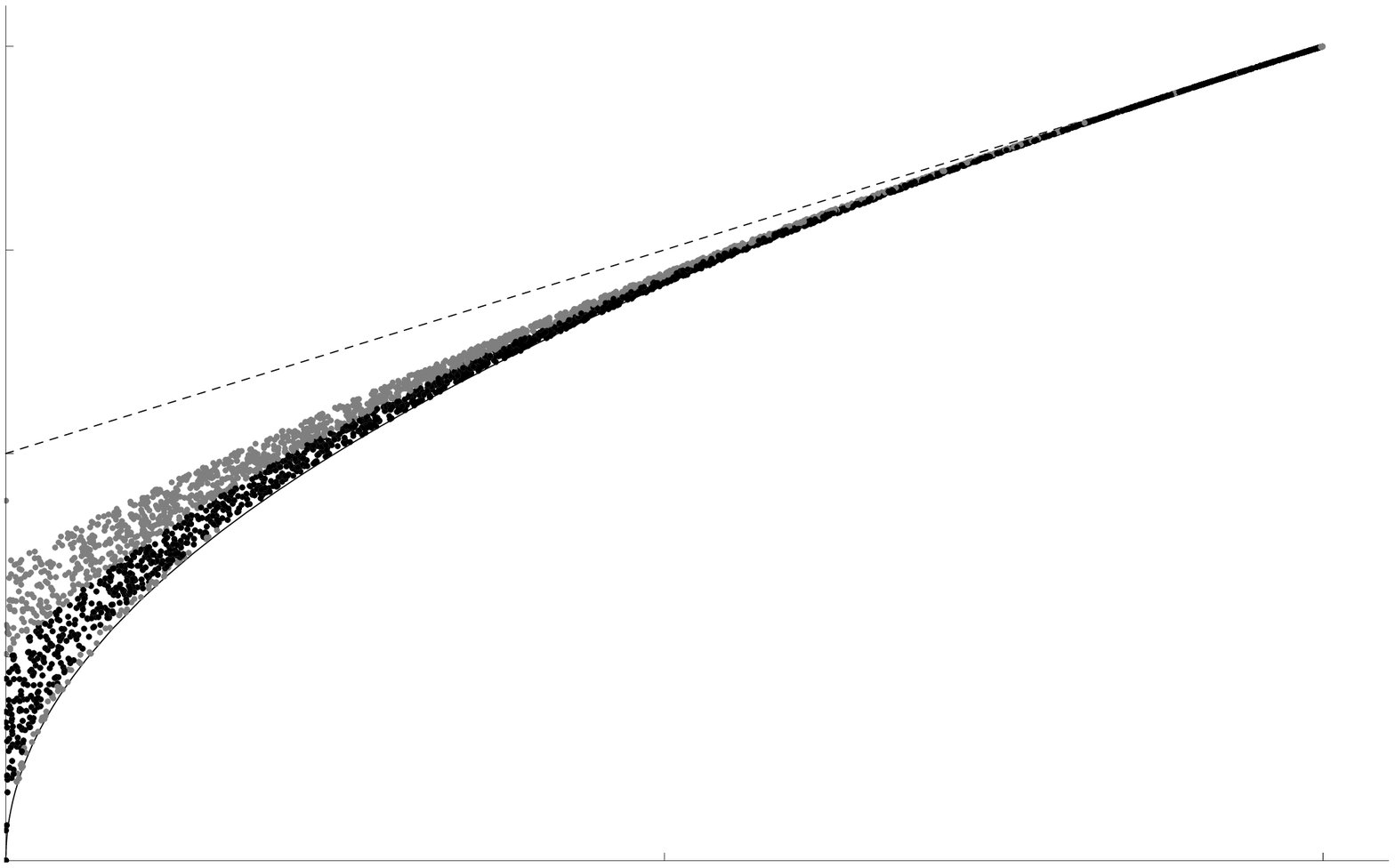}\hspace{1cm}
  \includegraphics[height=5cm,width=6.8cm]{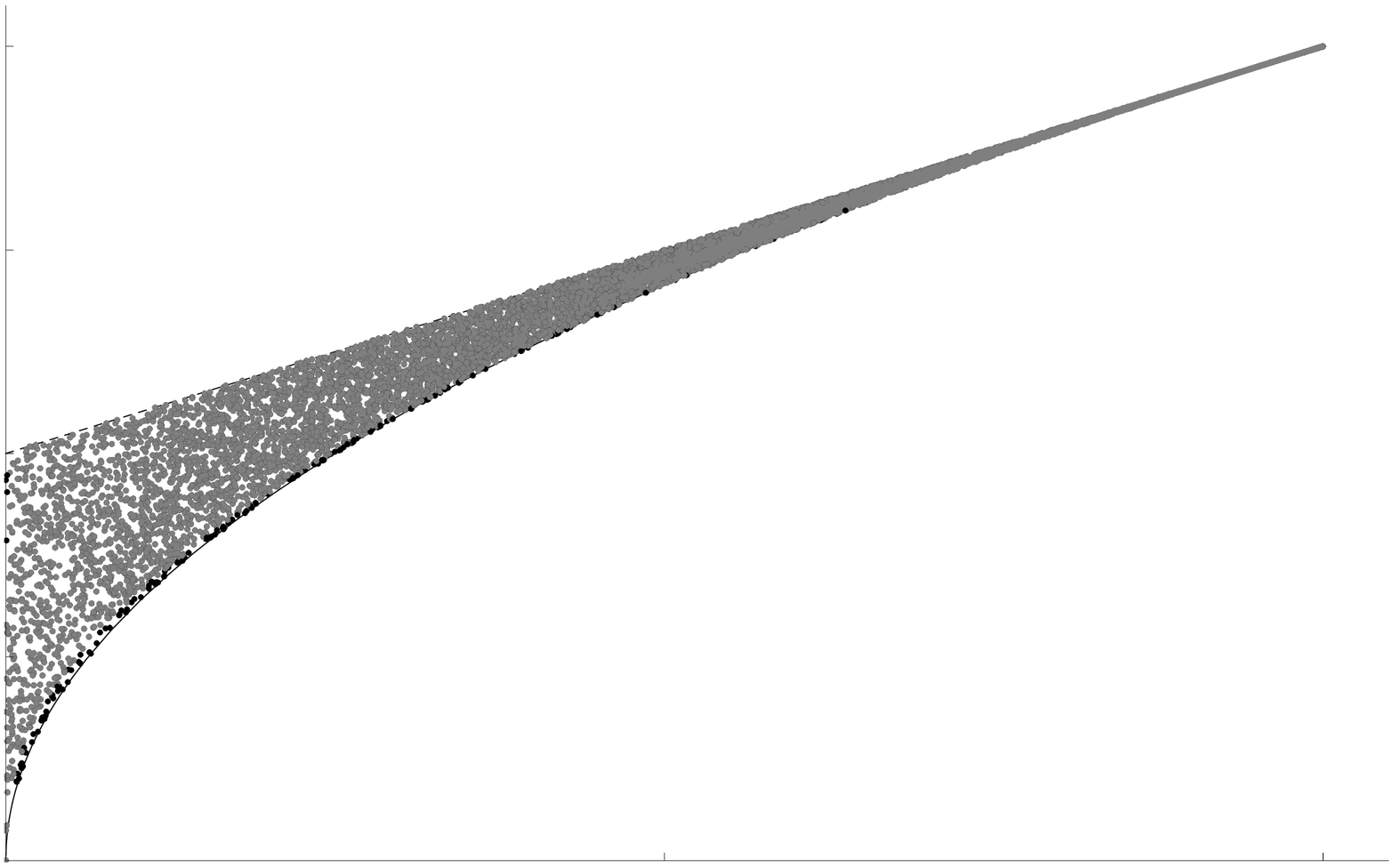}
\begin{picture}(0,0)(0,0)
\begin{tiny}
\put(-38,-2){$0.5$}
\put(-6,-2){$1$}
\put(-74,15){$0.5$}
\put(-72,23){$1$}
\put(-74,34){$1.5$}
\put(-72,45){$2$}
\put(-118,-2){$0.5$}
\put(-85,-2){$1$}
\put(-153,15){$0.5$}
\put(-152,23){$1$}
\put(-153,34){$1.5$}
\put(-152,45){$2$}
\end{tiny}
\begin{small}
\put(-70,51){$\lambda$}
\put(0,1){$\nu$}
\put(-149,51){$\lambda$}
\put(-80,1){$\nu$}
\end{small}
\end{picture}
\vspace{0.2cm}
\caption{\small Scatter plots in parameter phase space $(\nu,\lambda)$ for the doubly spinning black ring. 
The plot on the left shows the regions (10,000 points) where the heat capacities are negative, 
$C_{\Omega}<0$ (\textit{gray}) and $ C_{J}<0$ (\textit{black}). The regions where the compressibility 
$\epsilon_{\phi\phi}$ (\textit{gray}) and the $\det[\epsilon]$ (\textit{black}) are negative cover the entire 
parameter space (plot on the right) implying the local thermal instability of the doubly spinning 
black ring. The region in the parameter space is bounded: $0 \le \nu<1$ and $\lambda$ by the functions 
$1+\nu$ and $2\sqrt{\nu}$, shown as the \textit{dashed} and \textit{solid} lines respectively.}\label{fig:negzones}
\end{figure}

We consider first the specific heat at constant angular velocities 
\begin{eqnarray}
C_\Omega\equiv T\left(\frac{\partial S}{\partial T}\right)_{\Omega_{\phi},\Omega_{\psi}}
\end{eqnarray}
The analytic form of this quantity is too complicated to be written down here. 
Instead, we show on the left hand side in Fig. \ref{fig:negzones} a scatter 
region in the parameter space of the doubly spinning black ring where this 
heat capacity $ C_\Omega$ is negative (\textit{gray} -- 10,000 points). 
Note also that the parameters in the solution (\ref{eq:parameterrange}) are 
constrained and represented as a dashed line for $\lambda=1+\nu$ and solid 
line for the extremal black ring with $\lambda=2\sqrt{\nu}$.\\

In a similar way, we explore the region where the specific heat at
constant angular momentum
\begin{eqnarray}
\label{relcj}
C_J\equiv T\left(\frac{\partial S}{\partial T}\right)_{J_{\phi},J_{\psi}} 
\end{eqnarray}
is negative. The region (in black) where $C_J<0$ is shown in the scatter plot 
on the left of Fig. \ref{fig:negzones}. 

We observe a region in the parameter space of the doubly spinning black hole 
where both specific heats, $C_{\Omega}$ and $C_{J}$, can be positive simultaneously. 
However, this condition is not sufficient to draw the conclusion of thermodynamic 
stability: one should also investigate the matrix of `isothermal moment of inertia'.

These response functions are defined as
\begin{eqnarray}
\epsilon_{a b} \equiv \left(\frac{\partial J_{a}}{\partial \Omega_{b}}\right)_{T,\Omega_{a\ne b}}
\end{eqnarray}
We observe in Fig. \ref{fig:negzones} that the spectrum of the matrix of isothermal 
moment of inertia, spec[$\epsilon_{ab}$], is nowhere positive definite in the parameter 
space. 

Since there is no overlap region in parameter space in which all the response functions of 
interest are positive definite, we conclude that the doubly spinning black 
ring is unstable in the grand canonical ensemble.

\section{Instabilities from thermodynamics}\label{Instabilities}

Many known stationary black holes in higher dimensions present a 
black string or, more general, black brane phase --- we will refer 
to it as the `membrane phase'. That is, as the angular momenta are 
sufficiently increased (the ultra-spinning regime), the behaviour of 
some black holes and black rings changes to that of extended black 
branes and strings.

In the next subsection , we deal with the ultra-spinning 
black holes. From the study of the Gibbs potential's Hessian, we show 
the existence and find the locus of the transition points to the 
membrane phase. We also argue that there is a subtle relation between 
the microcanonical and grand canonical ensembles that may be at 
the basis of some of the results for ultra-spinning black 
holes discussed recently in \cite{Dias:2009iu}.

The analysis can be extended to (doubly) spinning black rings \footnote{The doubly spinning solutions we consider are with the spins in orthogonal planes. Other black hole solutions, where the two spins are parallely oriented, were also studied \cite{Evslin:2008py}.}. These 
results are presented in Section \ref{subsec:BlackMembrane}.

To compare different examples we will make use of the {\it dimensionless} expressions for the temperature $t$, the spin $j$, and the angular velocity $\omega$ defined by
\begin{equation}\label{toj}
t^{D-3} = c_{t}\, GM T^{D-3}\,\qquad \omega^{D-3}=c_\omega \,GM\Omega^{D-3} \,,\qquad j^{D-3}=c_j\,\frac{J^{D-3}}{GM^{D-2}}
\end{equation}
where the numerical constants are
\begin{small}
\begin{equation}
c_{t} =\frac{2}{(D-2)} \frac{(4\pi)^{D-3}}{\Omega_{D-3}} \left(\frac{D-3}{D-4}\right)^\frac{D-3}{2},\;
c_\omega=\frac{16}{(D-2)}\frac{(D-3)^{\frac{D-3}{2}}}{\Omega_{D-3}},\;
c_j =\frac{\Omega_{D-3}}{2^{D+1}}\frac{(D-2)^{D-2}}{(D-3)^{\frac{D-3}{2}}}\,.
\end{equation}
\end{small}

\subsection{Ultra-spinning black holes}
\label{subsec:UltraspinningBH}

Due to the qualitative changing behaviour of black holes as the 
dimensions are increased, the authors of \cite{Emparan:2003sy} 
have argued that the ultra-spinning black holes --- those in $D\ge 6$ 
dimensions that can have arbitrary large angular momentum per unit 
mass \cite{Myers:1986un} --- become unstable. The transition of 
these black holes, from behaving like a spherical black hole to 
behaving like a black membrane as the spin grows, was established 
to be at the minimum of the temperature. From that point onwards, 
the temperature increases in a similar way as for the black brane 
temperature.\\
\indent The minimum of the temperature where the behavior of the 
singly spinning black hole changes is determined as \cite{Emparan:2003sy}
\begin{equation}
\label{RuppBH1}
\frac{a^2}{r_h^2}=\frac{D-3}{D-5}
\end{equation}
This result was also obtained by using a different method, namely 
finding the divergences of the `Ruppeiner curvature'
 \cite{Ruppeiner:1995zz}. It was shown in 
\cite{Aman:2005xk} that, for a singly spinning Myers-Perry 
black hole, this curvature\footnote{The Ruppeiner curvature is the 
scalar curvature of the Hessian matrix of the entropy.} blows-up 
exactly at the value (\ref{RuppBH1}) signaling a thermal instability 
of the system.\\
\indent A qualitative understanding of this fact is related to the observation 
that, as the spin becomes large, the event horizon spreads out in the 
plane of rotation: it becomes a higher dimensional `pancake' approaching 
the geometry of a black brane.\\
\indent The existence of the ultra-spinning limit resembling black branes 
has a remarkable consequence. Black branes were shown to be 
classically unstable \cite{Gregory:1993vy} so that the ultra-spinning 
black holes would inherit the Gregory-Laflamme instability. The threshold 
of the classical instabilities and the connection to the 
thermal instability as conjectured by \cite{Gubser:2000mm} (see, 
also, \cite{Reall:2001ag}) requires a linearized analysis of the 
perturbations of the black hole solutions.

However, the transition to a membrane-like phase of the rapidly 
spinning black holes can be established from the study of the thermodynamics 
of the system. The existence and location of the threshold of this 
regime is signaled by the minimum of the temperature and the 
maximum angular velocity as functions of the angular momentum. 

It was observed in \cite{Dias:2009iu} that, for ultra-spinning black 
holes, this is in tight correspondence with a vanishing 
eigenvalue of the Hessian of the Gibbs potential. A complete 
thermodynamic analysis, though, should be based  on the full 
Hessian of the thermodynamic potential rather than only a study of 
the determinant.\footnote{A spinodal is 
defined as a line separating the regions of stability and instability of 
a homogeneous system. It is important to emphasize that all spinodals are 
zero-determinant lines, but in general not all zero-determinant lines are 
spinodal.} We will see in the next subsection that the membrane phase 
of a doubly spinning ring is not signaled by a zero-eigenvalue 
of the Gibbs potential's Hessian.

For ultra-spinning black holes, there is a {\it direct} relation between 
(some response functions in) the microcanonical and grand canonical 
ensembles.\footnote{Different ensembles correspond to different physical 
conditions and so, in more general cases, one does not expect such 
a relation.} To see that, let us compare the expressions of two 
particular response functions in these two ensembles:
\begin{equation}
\label{response}
\left(\frac{\partial^2 S}{\partial J^2} \right)_M = 
-\frac{1}{T}\left(\frac{\partial \Omega}{\partial J} \right)_M +
\frac{\Omega}{T^2}\left(\frac{\partial T}{\partial J} \right)_M \,\,\,\,\,\,\,\,\,\,\,\,\,\,
\text{and} \,\,\,\,\,\,\,\,\,\,\,\,\,\, 
\left(\frac{\partial^2 G}{\partial \Omega^2} \right)_T = -
\left(\frac{\partial J}{\partial \Omega} \right)_T
\end{equation}
We have checked that in the particular case of the singly spinning 
black hole, indeed, these two response functions are inverse 
proportional at the particular point where the temperature has a minimum. 
Therefore, an inflexion point in the microcanonical ensemble 
corresponds to a divergence of the corresponding response function in 
the grand canonical ensemble. This may well be an explanation for the results 
obtained in \cite{Dias:2009iu}. Moreover, this point should not be considered 
as a sign for an instability or a new branch but a transition to an infinitesimally nearby solution 
along the same family of solutions. The numerical evidence of \cite{Dias:2009iu} 
supports this connection with the zero-mode perturbation of the solution.

We now examine the situation for a more general family of ultra-spinning 
Myers-Perry black holes with multiple spin parameters, $a_i$, where 
$i=1,2,...,N$ and $N=\left[(D-1)/2\right]$. The black hole is characterized 
by the mass parameter $\mu$ and the horizon radius $r_h$ (the largest root of)
\begin{equation}\label{HorRadius}
 \mu=\frac{1}{r^{1+\epsilon}_h}\prod^{N}_{i=1}(r_h^2+a_i^2)  \,,
\end{equation}
by which we can express the thermodynamics 
\begin{eqnarray}\label{thermoMP}
M=\frac{\Omega_{D-2}}{16\pi G_D}(D-2) \,\mu ,\qquad J_i=\frac{\Omega_{D-2}}{16\pi G_D}a_i\,\mu,\qquad \Omega_i=\frac{a_i}{r_h^2+a_i^2}\,,\nonumber\\
{\cal A}_H=\Omega_{D-2} \,\mu \,r_h ,\qquad T=\frac{1}{2\pi r_h}\left(\,r_h^2\sum^{N}_{i=1}\frac{1}{r_h^2+a_i^2}-\frac{1+\epsilon}{2}\right)\,,
\end{eqnarray}
where $\epsilon=mod_2 D $. A sufficient, but not necessary, condition for the existence of ultra-spinning 
black holes was given in \cite{Emparan:2003sy}. In even(odd) dimensions at least one(two) 
of the spins should be much smaller than the rest. The ultra-spinning 
regime is obtained in the limit
\begin{equation}\label{parameters}
0\le a_1, a_2, ..., a_k<< a_{k+1}, ...,a_N\rightarrow \infty
\end{equation}
where $N-1\ge k\ge1+\epsilon$. The 
generic limiting black brane metric whether static, with all 
finite angular momenta {$a_1,..,a_k$} vanishing, or spinning, with 
some {$a_1,..,a_k$} non-vanishing, is the product $S^{D-2(N-k+1)}\times \mathbb{R}^{2 (N-k)}$.

Our focus will be on the case in which the black hole has at 
least two large spins and we set the remaining angular momenta 
to zero. When the angular momenta are equal, $J_{k+1}=...=J_N=J$, 
the Ruppeiner curvature scalar blows up 
at\footnote{Note that $\frac{a^2}{r_h^2}=\frac{4 J^2}{S^2}$ 
and in the particular case when $k=k_{max}$ our results agree 
with those of \cite{Aman:2005xk}.}
\begin{equation}
\label{RuppBH3}
\frac{a^2}{r_h^2}=\frac{D-3}{2k-1-\epsilon} \,.
\end{equation}
According with the arguments in \cite{Ruppeiner:1995zz}, this signals 
a thermodynamic instability. However, the expected new phase should 
correspond to the black membrane phase of ultra-spinning black holes 
and not to a new branch of solutions. 

This is further supported by examining the eigenvalues of the Hessian 
of the Gibbs potential. Indeed, we find that the divergences of the 
Ruppeiner curvature pinpoint the zero of the determinant of the Gibbs 
potential's Hessian.\\
\indent Also, by studying the temperature
\begin{equation}\label{temperature}
T= \frac{(D-3)\left(1+\frac{n}{(D-3)}\frac{4J^2}{S^2}\right)}{4 S^{\frac{1}{D-2}}\left(1+\frac{4 J^2}{S^2}\right)^{\frac{D-1+n}{2(D-2)}}}\,,\qquad n=2k-1-\epsilon
\end{equation}
we find that the temperature has a minimum at exactly 
(\ref{RuppBH3}), while the angular velocity $\Omega$ 
reaches its maximum value. Therefore for these more general ultra-spinning black holes, similarly to the singly spinning situation discussed in \cite{Emparan:2003sy}, once the minimum is reached the temperature 
increases and the angular velocity decreases signaling a transition to a 
membrane phase. This conduct is shown in figure \ref{fig:tvsj} ($I$ points on the solid thin line) for the particular case of $D=7$, $k=2$ so $j_1=j_2\equiv j_{\phi}$ and $j_3=0$.

Another case of interest is the ultra-spinning black holes that resemble 
spinning black branes, when some of the slower spins are non-zero $a_1,...,a_k\ne 0$.
It is not our goal to make a detailed analysis of this case here. Nevertheless, 
in all the cases where the non-vanishing spins are set equal, we find divergences 
of the Ruppeiner scalar curvature which could help to detect the 
threshold of their membrane phase.

\subsection{Membrane phase of black rings}
\label{subsec:BlackMembrane}

\begin{figure}
\centering
\includegraphics[height=6cm,width=7.5cm]{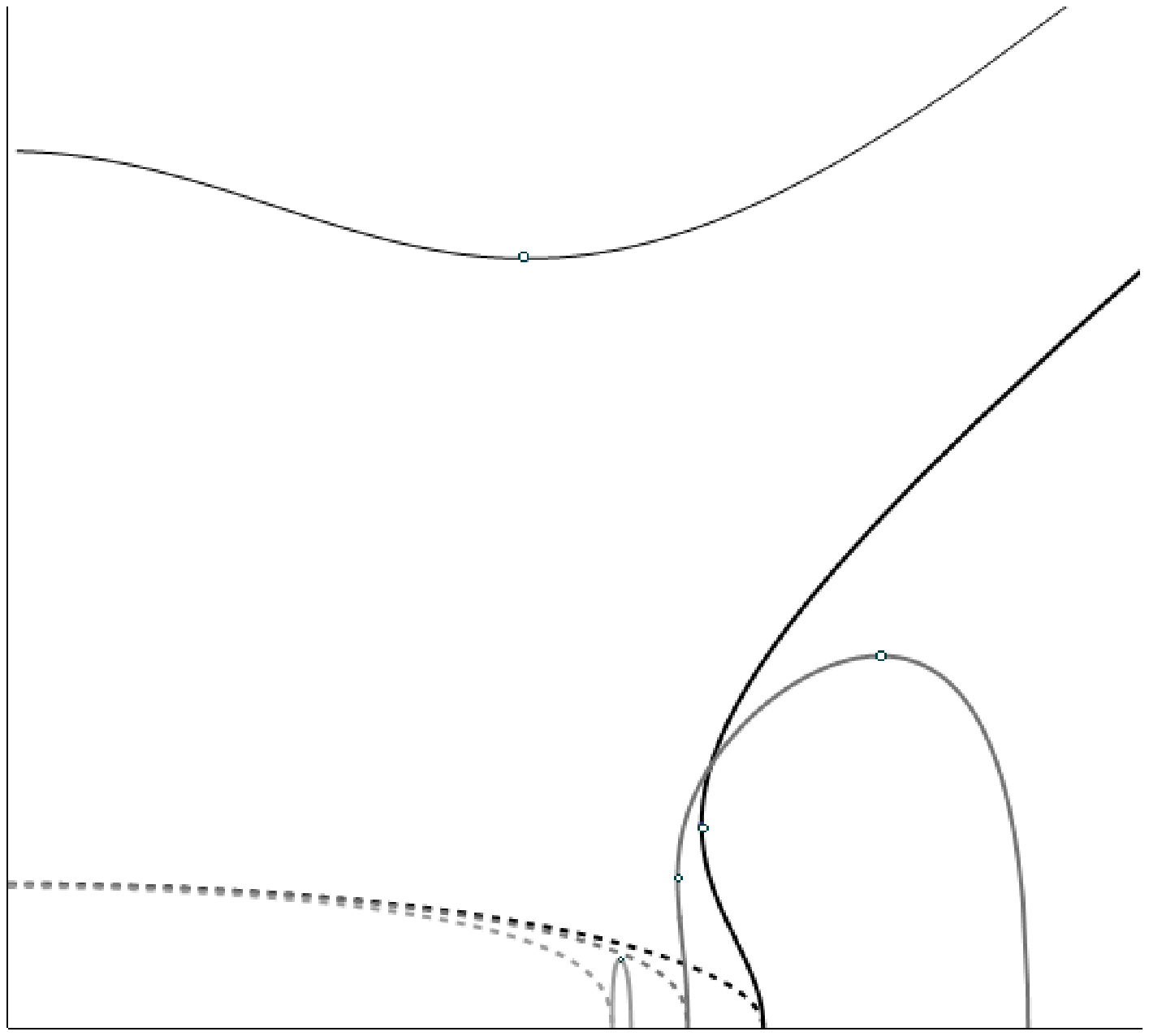}\hspace{0.5cm}
  \includegraphics[height=6cm,width=7.5cm]{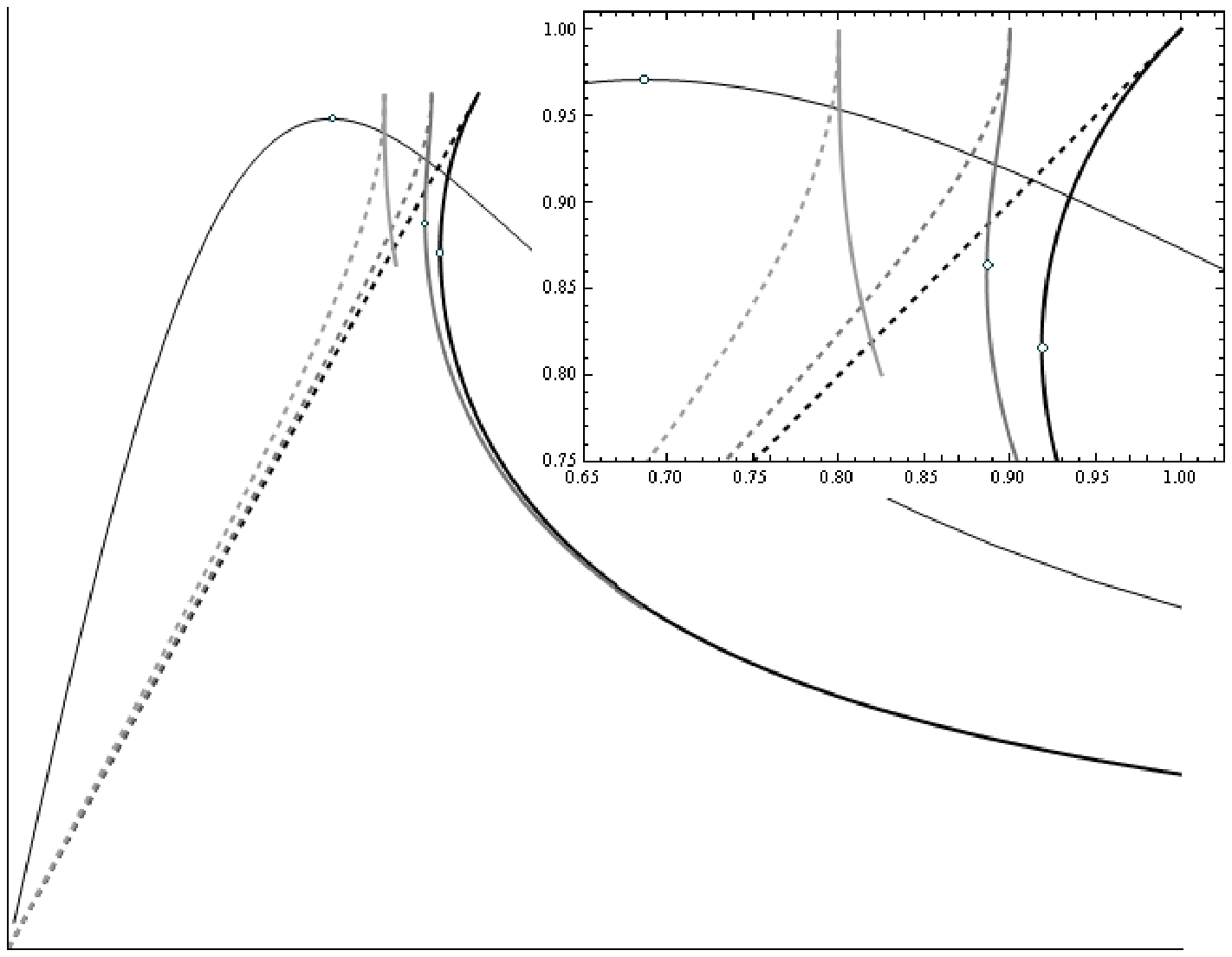}
\begin{picture}(0,0)(0,0)
\begin{tiny}
\put(-82,0){$j_{\phi}$}
\put(-157.5,61){$t$}
\put(-124.5,41){$I$}
\put(-111.5,12.2){$II$}
\put(-118,9){$III$}
\put(-101,23.5){$IV$}
\put(-118.2,5){$V$}
\put(-3,0){$j_{\phi}$}
\put(-77,61){$\omega_{\phi}$}
\put(-57,53.5){$I$}
\put(-49,43){$II$}
\put(-37.5,55.5){$I$}
\put(-12.5,37.5){$II$}
\put(-20.5,43){$III$}
\end{tiny}
\end{picture}
\vspace{0.1cm}
\caption{\small Plots of the temperature (left hand side) and 
angular velocity (right hand side)
as functions of the angular 
momentum $j_{\phi}$, for a fixed mass, for different 
black objects. These 
include the singly spinning Myers-Perry black holes in five dimensions 
of space-time (\textit{black dashed line}) and its seven dimensional 
cousin (\textit{solid thin line}). The singly (\textit{solid thick}) and 
doubly spinning black ring (\textit{light gray}) for different values 
of angular momenta (right towards left) $j_{\psi}=0.1,0.2$ on the $S^2$, 
are also shown here. For this same values of the second angular momenta the five-dimensional doubly spinning black holes are represented by the {\it dashed light gray} lines.}
\label{fig:tvsj}
\end{figure}

Other solutions, e.g. the black ring with one angular momentum, 
also exhibit an ultra-spinning behavior. The black ring, which
is characterized by the radii $r_0$ and $R$ of the spheres $S^{D-3}$ 
and $S^1$, respectively, becomes \textit{thin} in this limit 
(when $r_0<<R$).\footnote{This \textit{thin} regime was essential 
to find perturbatively the higher dimensional black rings cousins. 
Moreover, a generalization of this construction 
to black branes led to the construction of blackfolds 
\cite{Emparan:2007wm, Emparan:2009cs}.} 

Since the final expressions for the response functions are very 
complicated for the doubly spinning ring (see Appendix B), we prefer 
to present the `conjugacy diagram' of the angular velocity versus the 
angular momentum and the plot of the temperature as a function of 
the angular momentum, both for a fixed mass (see Fig. \ref{fig:tvsj}). 

For the singly spinning black ring, an analysis of the temperature 
as a function of the angular momentum was presented in 
\cite{Arcioni:2004ww}. In this case (solid thick line in the 
plot on the left hand side of Fig. $2$), 
the temperature does not have a minimum, but there 
exists a turning point.\footnote{We would like to point out that 
in (Sherk-Schwarz-)Anti-de Sitter, there is also a turning point 
\cite{Lahiri:2007ae}.} In our analysis, the turning 
point $II$ for the black ring plays a similar role as the minimum 
of the temperature for the 
black hole. That is, it signals a change in the thermodynamic 
behaviour of the black ring. In fact, it is the starting point 
of the ultra-spinning regime where the black ring can be 
approximated by a boosted black string.

Using the Poincar\'e `turning point' method, this special 
point was carefully studied in \cite{Arcioni:2004ww}. In particular, 
they found a divergence of the Ruppeiner curvature. In the conjugacy 
diagram (on the right) there is also a turning point $II$ at the same minimum 
value of the angular momentum $j_{\phi}$.

The question is then if there still is a relation between the microcanonical 
and grand canonical ensembles in this case. We have explicitly checked, using 
the results of \cite{Astefanesei:2009wi}, that one of the eigenvalues 
of the Hessian of the Gibbs potential is zero at this specific point $II$ 
while the second eigenvalue never changes its sign. Therefore, we conclude 
that the turning point is the onset of the ultra-spinning black string phase.\\

A far more richer structure is found for the doubly spinning black ring. 
The angular momentum on $S^2$ is bounded as $j_{\psi}\in [0,1/4]$
and for a specific $j_{\psi}$ the black ring can always be extremal 
(in the limit $\lambda\rightarrow 2\sqrt{\nu}$ as shown in \cite{Elvang:2007hs}). 
But besides extremality, according to how large $j_{\psi}$ is, the behaviour 
of the doubly spinning black ring changes. There are two distinctive 
regions in the microcanonical ensemble. On one hand, for $0\le j_{\psi}<1/5$, 
there are phases with the characteristic cusp for black rings with the two 
(fat and thin) branches. But on the other hand, for $1/5\le j_{\psi} \le 1/4$, 
the fat black ring branch disappears and so there are no cusps. As we will 
discuss in what follows, this will become relevant to understand the physics 
and regimes of the doubly spinning black ring.

To explore how the physics of the black rings at fixed mass is modified 
as we turn on the angular momenta along the $S^2$, $j_{\psi}$, we will 
study the temperature and angular velocity as functions of the $S^1$ 
angular momentum, $j_{\phi}$, for different fixed values of $j_{\psi}$.\\

\textbf{Doubly spinning black rings with $0\le j_{\psi}<1/5$}\\
In this case the situation is similar, to some extent, to what we found before for the 
singly spinning black ring. In this range, as we turn on $j_{\psi}$, there also 
are turning points with tangents of infinite slope signaling the onset 
of the black membrane phase that coincide with the cusps (in parameter space), 
namely at $\lambda=-(1/4)(1+\nu-(9+\nu)^{1/2}(1+9\nu)^{1/2})$. Figure 
\ref{fig:tvsj} shows this change in behavior explicitly. 
On the left (light gray curve) in the t vs. $j_{\phi}$ diagram, we observe 
the turning point $III$ that corresponds to the minimum  value of $j_{\phi}$ angular momentum. 
In the conjugacy diagram $\omega$ vs. $j_{\phi}$ (also the light gray curve), 
the corresponding point ($III$) is a turning point.\\
\indent But an interesting difference with the single spin 
black ring occurs for a large enough $S^1$ angular momentum. 
The temperature of the black ring in its membrane phase increases while the area of the 
event horizon decreases up to a point where the spin-spin interaction 
is large enough making a turn to abruptly become extremal 
with zero temperature. This is the maximum critical value labeled 
$IV$ in figure \ref{fig:tvsj}. Therefore, the black membrane phase exists between 
points $III$ and $IV$.\\ 
\indent Finally, as for the singly spinning black ring, we computed the eigenvalues of the Hess[G] and found that none of them are zero the relevant turning points.\\

\textbf{Doubly spinning black rings with $1/5\le j_{\psi} \le 1/4$}\\
This black rings (lightest gray line in figure \ref{fig:tvsj}) 
with larger $S^2$ angular momentum show no turning points 
and therefore have no membrane phase. This mimics its behavior 
in the phase diagram of microcanonical ensemble where the cusp 
and fat branch of these doubly spinning black ring disappear \cite{Elvang:2007hs}.
The lack of a fat black ring branch seems to coincide with the lack of a 
black membrane phase. Therefore such solutions would never be captured with 
long distance effective approaches \cite{Emparan:2007wm, Emparan:2009cs}.\\ 
\indent Note that for certain fixed $j_{\psi} \in [1/5, 1/4]$ the temperature grows, reaching a maximum at $V$ 
and rapidly decreasing to zero to become extremal. It would be interesting to explore the physical meaning of these points which we observe to correspond to an inflection point $\partial^2 S/\partial^2 J \,= 0$ for fixed mass.\\

In summary, we have found examples for which the zero eigenvalues of the Hessian of the 
Gibbs potential can also be turning points (with tangents of infinite slope) 
and not just critical points as for the Myers-Perry black holes.
Other less symmetric solutions, such as the doubly spinning black ring, do not show this connection between the zeros of the Hess[G] and the onset of the black membrane phase.
Moreover, we showed that certain doubly spinning black ring, those with $1/5\le j_{\psi} \le 1/4$, have no membrane 
phase. Therefore, these particular solutions fall into 
the same category as other black holes with no membrane phase as the 
four dimensional Kerr black hole and the five dimensional Myers-Perry 
black hole.

\section{Discussion}\label{discussion}

In this paper we have analyzed in detail the thermodynamic stability of 
neutral doubly spinning black objects.  
We have analytically computed the response functions and presented strong 
evidence showing that the doubly spinning black ring is thermodynamically unstable. That is, 
 there is no region in the parameter space in which all the response 
functions are positive definite.\\
\indent We have provided an explanation of why the microcanonical and grand 
canonical ensembles for ultra-spinning black holes are related in a very 
specific way. 
An inflexion point in the microcanonical ensemble corresponds to a 
divergence of the corresponding response function in the grand canonical ensemble.
We will comment on the validity of this argument and on the significance 
of these results in the last part of this section.\\
\indent The onset of a membrane phase of different 
doubly spinning black objects was identified. We have 
found that the onset of the black membrane phase 
for all black holes that we have studied (except for the less symmetric doubly spinning black ring) 
 is characterized by at 
least one zero eigenvalue of the Hessian of the Gibbs potential. A tight relation with the
 classical perturbations, where a transition to an infinitesimally 
nearby solution of the same family branch happens, is expected
in the cases where the connection between membrane and zero-eigenvalue of Hess[G]
exists. The numerical evidence of \cite{Dias:2009iu, Dias:2010eu} supports this connection 
precisely with the zero-mode perturbation of the solution.\\
\indent We now discuss the thermodynamics of doubly spinning 
black rings in the grand canonical ensemble. 
An analysis of this solution in the microcanonical ensemble was presented 
in \cite{Elvang:2007hs}. In general, for black holes, the entropy is 
used to obtain the phase diagrams in the microcanonical ensemble while 
the mass is kept fixed. However, in general relativity it 
makes more sense to use the total energy instead of the entropy. The 
reason is that this would require appropriate boundary conditions. \\
\indent It is important to emphasize that it is not known how the background 
subtraction method can be applied to black rings,
because it is not clear how to choose the background solution.  
We therefore used the counterterm 
method to compute the action and the grand 
thermodynamic potential (which is a Legendre transform of the energy). 
Since the expressions of the response functions are too complicated for analytical treatment
(see Appendix  B), we have plotted the regions in the parameter space where they are positive.\\
\indent Let us know compare some of our results with other well understood 
examples --- at this point we are interested just in the thermodynamic 
instabilities, not in the relation with dynamical ones.
The fact that the Schwarzschild black hole has a negative heat capacity means that 
the thermodynamic ensemble is dominated by diffuse radiation 
states rather than black holes states but is classically dynamically stable at the 
linearized level. When adding angular momentum the situation changes and the heat 
capacity can be positive for a large enough angular momentum. However, 
this condition is not enough to conclude that the system is
thermodynamically stable. The stability also implies that when 
angular momentum is added to the system the angular velocity goes up.

\begin{figure}
\centering
  \includegraphics[height=6cm]{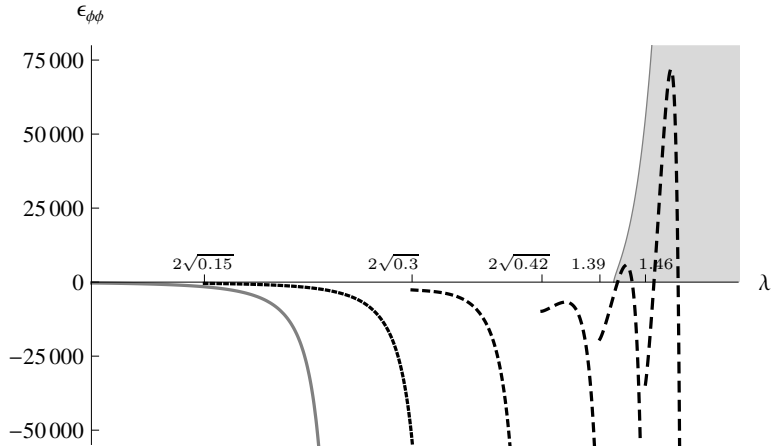}
\begin{picture}(0,0)(0,0)
\begin{tiny}
\put(-81,23.5){$2\sqrt{0.15}$}
\put(-55,23.5){$2\sqrt{0.3}$}
\put(-39,23.5){$2\sqrt{0.42}$}
\put(-28,23.5){$1.39$}
\put(-19,23.5){$1.46$}
\end{tiny}
\end{picture}
\vspace{0.2cm}
\caption{\small Plot of the response function $\epsilon_{\phi\phi}$ as a function of $\lambda$. 
The \textit{gray} curve corresponds to the compressibility of the singly spinning black ring, 
namely $\nu=0$. As the angular momentum along the $S^2$ is increased (\textit{dashed line} 
left towards right) the isothermal moment of inertia for different values of 
$\nu= 0.15,0.3,0.42,0.48,0.53 $ changes and becomes positive for values of 
$\nu>0.46690042$ and $\lambda>1.40685$ (shaded \textit{gray region}). }\label{fig:epsilonanalisis}
\end{figure}

\indent For a black ring with one angular momentum the heat capacity can be 
positive, but the momentum of inertia is always negative \cite{Astefanesei:2009wi}. Therefore, 
the singly spinning black ring is thermodynamically unstable. As in 
the case of one angular momentum, the heat capacity of a doubly 
spinning black ring can be positive in some region of the parameter 
space. However, there is a key difference when the second angular momentum 
is turned on. That is, the component of the momentum of inertia associated 
to $S^1$ of the black ring can become positive --- this is explicitly 
shown in Figure \ref{fig:epsilonanalisis}. \\
\indent Since there are two angular momenta one should also investigate the 
effect of coupled `angular' inhomogeneities. A careful study of the 
determinant of the momentum of inertia matrix shows that there is no 
region in the parameter space with the desired properties and 
so the doubly spinning black ring is also thermodynamically unstable.

\begin{figure}
\centering
  \includegraphics[height=5cm,width=6.8cm]{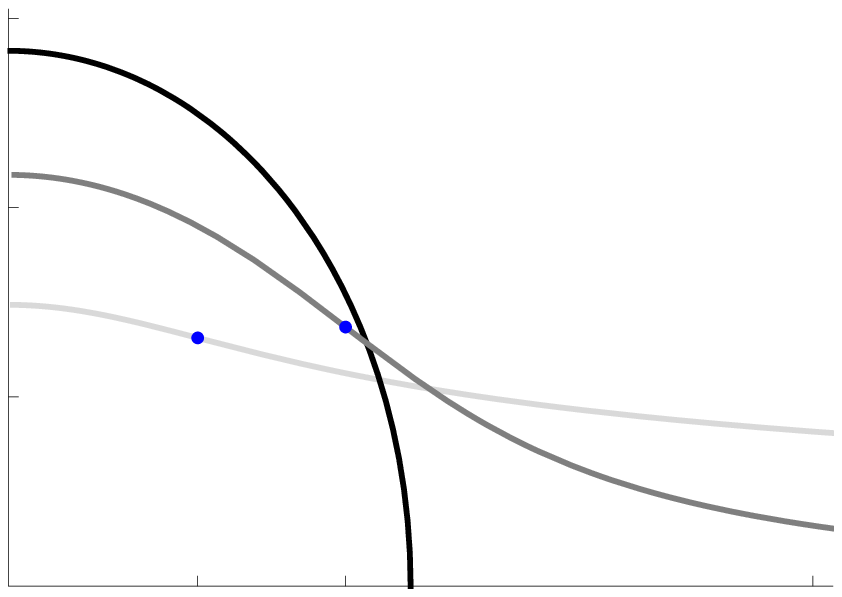}\hspace{1cm}
 \includegraphics[height=5cm,width=6.8cm]{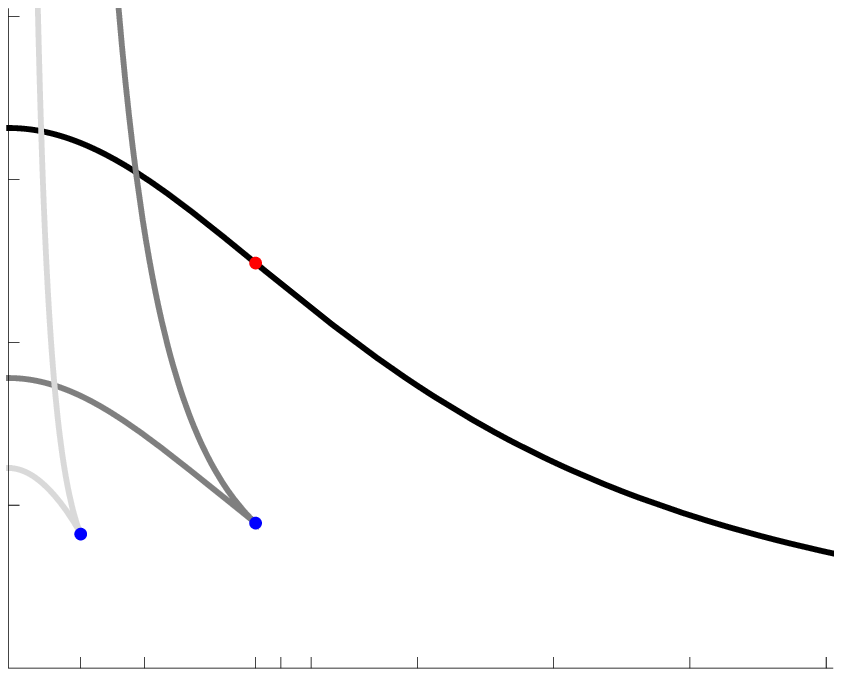}
\begin{picture}(0,0)(0,0)
\begin{small}
\put(-149,52){$s$}
\put(-79,2){$j$}
\put(1,2){$\omega$}
\put(-69,53){$g$}
\end{small}

\begin{tiny}
\put(-134,20){$A$}
\put(-122,21){$B$}
\put(-63,9){$A$}
\put(-49,10){$B$}
\put(-50,28){$C$}
\end{tiny}
\end{picture}
\vspace{0.2cm}
\caption{\small The microcanonical phase diagram - entropy, s, as a function of the angular momentum, j - for fixed mass 
(on the \textit{left}) of the singly spinning Myers-Perry black hole in $D=5,6,10$ dimensions of space time 
(\textit{black}, \textit{gray} and \textit{lightgray} lines respectively). The grand canonical phase diagram 
(on the \textit{right}) - Gibbs potential, g, as a function of the angular velocity, $\omega$ - for fixed 
temperature of the black hole. The points $A,B$ correspond to $a/r_h=\sqrt{3},\sqrt{\frac{7}{5}}$ for six 
and ten dimensions, and in general for $D>5$ the points are $a/r_h=\sqrt{(D-3)/(D-5)}$; the change in convexity 
of the entropy corresponds to a blowing up of the convexity of the Gibbs potential. The point C instead, at 
$a/r_h=1/\sqrt{3}$ and where the Gibbs potential has a change in convexity, has no analog in the microcanonical 
scheme.}\label{fig:microvsgrandcanonical}
\end{figure}

We would like now to discuss in more detail some of the results for 
ultra-spinning black holes presented in Section $4$.  In Fig. $\ref{fig:microvsgrandcanonical}$, we 
show the points (A,B) in the grand canonical ensemble  that correspond 
to inflexion points (A,B) in the microcanonical ensemble. This can be 
quantitatively understood by comparing the particular response 
functions in eq. (\ref{response}) at a very special point in the parameter 
space, namely where the temperature has a minimum.\\
\indent We have explained that this argument applies to this particular 
case but not in general. A counterexample is the $5$-dimensional black 
hole with one angular momentum. In this case there is no relation 
between ensembles in the sense that there is no special point in 
the microcanonical ensemble which corresponds to the inflexion 
point ($C$) in the grand canonical ensemble. 
Moreover we have checked and there 
are no points where an eigenvalue of the Hessian of the Gibbs potential 
vanishes. Therefore, it should not be considered 
as a sign for a membrane phase --- most probably, it is similar 
with the Schwarschild black hole example for which the thermodynamical 
instability is not related to a dynamical one.

One can also consider the ultra-spinning black holes with some 
of the finite angular momenta non-zero. In odd(even) spacetime 
dimensions, the metric of an ultra-spinning black hole with all 
but two(one) of the spins finite and non-zero will reduce to that 
of a spinning black brane. As we have shown in Section 2, the 
counterterm method can also be applied to spinning black branes 
and the results are similar to the ones for the `seed' spinning black hole solution. We 
have computed the \textit{renormalized} action to find the Gibbs 
potential and we expect similar thermal instabilities as for the 
corresponding black holes. As we already emphasized, though, in 
all these examples we expect that the thermodynamic instabilities 
do not signal a dynamical instability or a new branch, but rather 
a transition to an infinitesimally nearby solution along the same 
family of solutions \cite{Dias:2009iu}. \\
\indent It is remarkable that our study of thermodynamic instabilities provides 
to some extent information 
about the zero-mode of the `Gregory-Laflamme instability' and it may well 
be the starting point for a more detailed study of dynamical 
instabilities. The extension of the dynamical stability studies 
to spinning black branes has not been yet developed. The analytic 
theory of perturbations is much more involved. However, we expect that the spinning black branes suffer from 
similar instabilities as the static ones. We hope that the observations made in this paper will be 
useful in future investigations of the perturbations of 
higher-dimensional spinning black rings, black holes and 
spinning black branes.

\section*{Acknowledgments}
We would like to thank Ido Adam for interesting conversations. DA and 
MJR would also like to thank Robb Mann and Cristian Stelea for collaboration 
on related projects and valuable discussions.

\appendix

\section{Temperature and angular velocities}\label{Ap:Temperature}
Consider a general stationary $5$-dimensional metric that
corresponds to a black object with two angular 
momenta\footnote{A similar analysis for one angular momentum can 
be found in \cite{ Astefanesei:2009wi,Astefanesei:2007bf}.} :
\begin{eqnarray}
\label{generalmetric}
ds^2 &=& g_{tt}(\vec{x}) dt^2+2 g_{t\phi}(\vec{x}) dt d\phi+2 g_{t\psi}(\vec{x}) dt d\psi+ 
g_{\phi\phi}(\vec{x})\,d\phi^2
\nonumber\\
&+& 2 g_{\phi\psi}(\vec{x})\,d\phi d\psi+ g_{\psi\psi}(\vec{x})\,d\psi^2+g_{\alpha\beta}(\vec{x})\,dx^\alpha dx^\beta
\end{eqnarray}
$\partial_t$, $\partial_\phi$, and $\partial_\psi$ are Killing 
vectors. Rewrite the metric in the ADM form 
\begin{eqnarray}
ds^2=-N^2 dt^2+\gamma_{ij}(dx^i+N^i\,dt)(dx^j+N^j\,dt)
\end{eqnarray}
with lapse function 
\begin{eqnarray}
N^2=-g_{tt}+g_{\phi\phi}\,(N^{\phi})^2+g_{\psi\psi}\, (N^{\psi})^2+2\, g_{\phi\psi}\, N^{\phi}N^{\psi}
\end{eqnarray}
and shift vector 
\begin{eqnarray}
N^{\phi}=\frac{g_{t\psi}g_{\phi\psi}-g_{\psi\psi}g_{t\phi}}{g_{\phi\psi}^2-g_{\phi\phi}g_{\psi\psi}}\,,
\qquad N^{\psi}=\frac{g_{t\phi}g_{\phi\psi}-g_{\phi\phi}g_{t\psi}}{g_{\phi\psi}^2-g_{\phi\phi}g_{\psi\psi}}
\end{eqnarray}

The event horizon is obtained for
\begin{eqnarray}
\label{hor}
N^2=0
\end{eqnarray}
In other words, it is a Killing horizon of 
$\partial_t + \Omega_\phi \, \partial_\phi + \Omega_\psi\,\partial_\psi$, 
where $\Omega_\phi$ and $\Omega_\psi$ are the angular velocities defined
as the shift vectors at the horizon:
\begin{eqnarray}
\label{angvel}
\Omega_{\phi}\left.=-N^{\phi}\right|_H\,,\qquad\Omega_{\psi}=\left.-N^{\psi}\right|_H
\end{eqnarray}
Black holes are thermodynamic objects: the causal structure of 
spacetime can influence the physics of a quantum field. The vacuum 
fluctuations near the event horizon cause the black hole to emit 
particles with a thermal spectrum. The Euclidean regularity at the 
horizon is equivalent to the condition that the black hole is in 
thermodynamical equilibrium.

By a straightforward computation one can eliminate the conical 
singularity in the Euclidean section, $(it,r)$, to obtain the 
periodicity of the Euclidean time. In this way, one obtains the 
following expression for the temperature of the black hole:
\begin{eqnarray}
\label{temp}
T=\left.\frac{(N^2)'}{4\pi\sqrt{g_{rr}\, N^2}}\right|_H
\end{eqnarray}
We have used these definitions to compute the corresponding 
physical quantities of the doubly spinning ring.

\section{Conditions for thermodynamic stability } \label{Ap:Stability}

In this appendix, we present the conditions for the thermodynamic
stability and we also give some useful explicit expressions for the 
response functions used in Section 4 --- we follow closely \cite{callen}.

For simplicity, let us start with a black hole with one angular momentum. 
We are interested in the thermodynamic potentials: the energy and 
its Legendre transforms. 

The basic extremum principle of thermodynamics (for the entropy $S$)
implies both that $dS=0$ and that $d^2S<0$. The second condition determines 
the stability of predicted equilibrium states. The stability criterion in 
energy representation requires that an equilibrium state at 
fixed $S$ and $J$ is a state of minimum energy, namely a minimum of $E[S,J]$. 
The local stability conditions ensure that inhomogeneities of either $S$ and $J$ separately 
\begin{equation}\label{e1}
\left(\frac{\partial^2 E}{\partial S^2}\right)_{J}=\left(\frac{\partial T}{\partial S}\right)_{J}\geq 0, \hspace{1cm} \left(\frac{\partial^2 E}
{\partial J^2}\right)_{S}=\left(\frac{\partial \Omega}
{\partial J}\right)_{S}\geq 0
\end{equation}
and {\it also} that a coupled inhomogeneity of $S$ 
and $J$ together 
\begin{equation}
\det({\rm Hess}(E))=\frac{\partial^2 E}{\partial S^2}\,\frac{\partial^2 E}{\partial J^2}-\left(\frac{\partial^2 E}{\partial S \partial J}\right)^2\geq 0
\label{e2}
\end{equation}
do not decrease the energy.

In more generality the stability criterion states that the thermodynamic 
potentials are {\it convex} functions of their {\it extensive} 
variables and {\it concave} functions of their {\it intensive} 
variables (see, e.g., \cite{callen}).

For a grand canonical ensemble defined at fixed temperature 
$T$ and angular velocities $\Omega_a$ (intensive variables) 
the associated potential, the Gibbs free energy, satisfies the following relations:
\begin{eqnarray}
\label{G}
G[T,\Omega] = E-TS-\Omega J\,\, , \,\,\,\,\,\,\,\,\,
dG = -SdT-Jd\Omega
\end{eqnarray}
In this case, the local stability conditions following from the convexity of 
the Gibbs function  yield
\begin{equation}
\left(\frac{\partial^2 G}{\partial T^2}\right)_{\Omega}=-\left(\frac{\partial S}{\partial T}\right)_{\Omega}\leq 0, \hspace{1cm} \left(\frac{\partial^2 G}
{\partial \Omega^2}\right)_{T}=-\left(\frac{\partial J}
{\partial \Omega}\right)_{T}\leq 0
\end{equation}
and 
\begin{equation}
\det({\rm Hess}(G))= \frac{\partial^2 G}{\partial T^2}\,\frac{\partial^2 G}{\partial \Omega^2}
-\left(\frac{\partial^2 G}{\partial T \partial \Omega}\right)^2\geq 0
\end{equation}
Equivalently, the heat capacities ($C_{\Omega}$, $C_J$) and the `isothermal 
moment of inertia' or 'compressibility' $\epsilon\equiv(\partial J/\partial \Omega)_T$ should be positive definite.

A generalization for two angular momenta is straightforward (see, also, 
\cite{Monteiro:2009tc}). The Hessian is a $3\times3$ matrix
\[{\rm Hess}(G) = (-1)\left(
\begin{array}{ccc}
        C_{\Omega}\,T^{-1}\,&\alpha_{b}\\
        \alpha_{a}\,\,&\epsilon_{ab}\\
\end{array} \right) \]
where the matrix components are 
$\alpha_{a}=\left(\frac{\partial J_{a}}{\partial T}\right)_{\Omega}$, $\epsilon_{ab}
=\left(\frac{\partial J_a}{\partial\Omega_b}\right)_T$, 
and the indices cover the angular directions $a,b=\phi,\psi$. 

Considering the relationship between the specific heats 
$C_{\Omega}=C_J+T\,(\epsilon^{-1})_{ab}\,\alpha^a\alpha^b$ 
it can be shown that a thermodynamically stable system is characterized by
positive heat capacities $C_\Omega>0$ and $C_J>0$ and, also, 
a positive definite matrix of isothermal momenta of inertia, i.e. 
${\rm spec}[\epsilon_{ab}]>0$.

The $(\phi\phi)$-component of the isothermal moment of inertia tensor is
\begin{eqnarray}
\nonumber
\epsilon_ {\phi\phi} &=& -\frac{4 k^4 \pi  \lambda  (1+\lambda +\nu )}{G_5 (\nu-1)^4
(1-\lambda +\nu )^2}\, \frac{F(\lambda,\nu)}{\sqrt{\lambda ^2-4 \nu }(\lambda ^2+
\lambda ^2\nu-8\nu) + \lambda ^3 (1+\nu)-2 \lambda  (4\nu+\nu^2-1)}
\end{eqnarray}
where
\begin{eqnarray}
\nonumber
F(\lambda,\nu)&=&\lambda ^6 (1+\nu )^2+\lambda ^5 (1+\nu )^2 (1+\sqrt{\lambda ^2-4
\nu }+\nu )
\nonumber\\
&&
+ \lambda ^4 (1+\nu )[4+\sqrt{\lambda ^2-4 \nu }+\nu  (2 (1+\sqrt{\lambda ^2-4 \nu})+\nu  (-24+\sqrt{\lambda ^2-4 \nu }+2 \nu) ) ]
\nonumber\\
&&
-16 \sqrt{\lambda ^2-4 \nu } + \nu  [1+\nu  (-14+\nu (18+(-14+\nu ) \nu ))]
\nonumber\\
&&
+2 \lambda ^3 (1+\nu ) [2+\sqrt{\lambda ^2-4 \nu
}+\nu  (-25+\sqrt{\lambda ^2-4 \nu }+\nu  (13-11 \sqrt{\lambda
^2-4 \nu }
\nonumber\\
&& 
+\nu  (-7+\sqrt{\lambda ^2-4 \nu }+\nu)))]+2 \lambda ^2 (1+\nu ) [2+\sqrt{\lambda
^2-4 \nu }+\nu  (-32-26 \sqrt{\lambda ^2-4 \nu }
\nonumber\\
&&
+\nu  (32+14
\sqrt{\lambda ^2-4 \nu }+\nu  (16-6 \sqrt{\lambda ^2-4 \nu
}+(-2+\sqrt{\lambda ^2-4 \nu }) \nu)))]
\nonumber\\
&&
-4 \lambda  [-1+\nu  (10+4\sqrt{\lambda ^2-4 \nu }+\nu  (-89-12 \sqrt{\lambda ^2-4 \nu }+\nu
(-12 (-6+\sqrt{\lambda ^2-4 \nu })
\nonumber\\
&&
+ \nu  (-23+4\sqrt{\lambda ^2-4 \nu }+(-2+\nu ) \nu))))]
\nonumber
\end{eqnarray}
The determinant is
\begin{eqnarray}
\epsilon &=& \frac {4 k^8 \pi ^2 \lambda ^2 (\lambda -\sqrt{\lambda ^2-4
\nu })  (\lambda +\sqrt{\lambda ^2-4 \nu })^2 \sqrt{\nu }
(1+\lambda +\nu )^5}{(G_5)^2 (-1+\nu )^4 [-\lambda
^2+(1+\nu )^2 ]^{3/2} [\nu  (-\lambda ^2+(1+\nu
)^2)]^{3/2}}\,\, \frac{G(\lambda, \nu)}{Z(\lambda, \nu)} 
\end{eqnarray}
where
\begin{eqnarray}
\nonumber
Z(\lambda, \nu) &=& 8 \sqrt{\lambda ^2-4 \nu } \nu -\lambda ^3
(1+\nu )-\lambda ^2 \sqrt{\lambda ^2-4 \nu } (1+\nu )+2 \lambda  (-1+4
\nu +\nu ^2)
\nonumber\\
G(\lambda, \nu)&=&\lambda ^5+7 \lambda ^4 (1+\nu )-\lambda ^2 (1+\nu
) [1+3 \sqrt{\lambda ^2-4 \nu }+(2-3 \sqrt{\lambda ^2-4 \nu
}) \nu +\nu ^2]
\nonumber\\
&&+
\lambda ^3 [5-6 \sqrt{\lambda ^2-4 \nu
}+(26+6 \sqrt{\lambda ^2-4 \nu }) \nu +5 \nu ^2]-8 \nu
(1+11 \nu +11 \nu ^2+\nu ^3)
\nonumber\\
&&
-\lambda [(8-27 \sqrt{\lambda ^2-4 \nu }) \nu +9 (16+3
\sqrt{\lambda ^2-4 \nu }) \nu ^2+(8+3 \sqrt{\lambda ^2-4 \nu
}) \nu ^3-3 \sqrt{\lambda ^2-4
\nu }]
\nonumber
\end{eqnarray}


\end{document}